\documentclass[aps,prb,twocolumn,showpacs,preprintnumbers,amsmath,amssymb,superscriptaddress]{revtex4-1}
\usepackage{graphicx}
\usepackage{dcolumn}
\usepackage{bm}
\usepackage{epstopdf}
\usepackage{textcomp}
\usepackage{mathtools}
\usepackage{float}
\usepackage{amsmath}

\begin{document}

\title{Complex magnetism and non-Fermi liquid state in the vicinity of the quantum critical point in the CeCo$_{1-x}$Fe$_x$Ge$_3$ series}

\author{P. Skokowski}
\email{przemyslaw.skokowski@ifmpan.poznan.pl}	
\affiliation{Institute of Molecular Physics, Polish Academy of Sciences\\
Smoluchowskiego 17, 60-179 Pozna{\'n}, Poland}

\author{K. Synoradzki}
\affiliation{Institute of Molecular Physics, Polish Academy of Sciences\\
Smoluchowskiego 17, 60-179 Pozna{\'n}, Poland}
\affiliation{Institute of Low Temperature and Structure Research, Polish Academy of Sciences,
Ok{\'o}lna 2, 50-422, Wroc{\l}aw}

\author{T. Toli{\'n}ski}
\affiliation{Institute of Molecular Physics, Polish Academy of Sciences\\
Smoluchowskiego 17, 60-179 Pozna{\'n}, Poland}	


\begin{abstract}
We report extensive studies on the CeCo$_{1-x}$Fe$_{x}$Ge$_3$ alloys, which show quantum critical point (QCP) due to damping the antiferromagnetic order in CeCoGe$_3$ down to $0 \ {\rm K}$ by doping with the paramagnetic CeFeGe$_3$ compound. The presence of QCP is confirmed by detecting the non-Fermi liquid behavior (NFL) using a wide range of the experimental methods: magnetic susceptibility, specific heat, electrical resistivity, magnetoresistance, and thermoelectric power. In the case of the thermoelectric power we find a clear enhancement of the Seebeck coefficient for $x$ around 0.6, i.e. in the neighborhood of QCP. Finally, the different complementary studies enabled construction of the complex magnetic phase diagram for the CeCo$_{1-x}$Fe$_{x}$Ge$_3$ system, including the energy scale imposed by the crystal electric field splitting of the Ce ground state. 
\end{abstract}


\maketitle

\section{\label{intr}Introduction}

Ce-based alloys and compounds as a prototype examples of strongly correlated electron systems, have been intensively investigated since their discovery due to their exotic properties and fascinating physics.\cite{weng_multiple_2016, steglich_foundations_2016, wirth_exploring_2016} The study of the anomalous physical properties close to the borderline between magnetically ordered and nonmagnetic ground states is of intense interest in contemporary condensed matter physics research. This point of instability achieved by non-thermal tuning parameters between two stable phases of matter is called the quantum critical point (QCP).\cite{doniach_kondo_1977} In the vicinity of the QCP usually non-Fermi liquid (NFL) behavior or unsual superconductivity appears. There are several methods to tune a system to the QCP with manipulating a control parameter $\delta$, most known are applying a hydrostatic pressure to the sample, using a strong magnetic field or a chemical substitution. The intense scientific work focused on the QCP was based on a few famous compounds and systems, such as CeCu$_{6-x}$Au$_x$,\cite{lohneysen_non-fermi-liquid_1994,schroder_onset_2000, rosch_mechanism_1997,schroder_scaling_1998,stockert_two-dimensional_1998, klein_signature_2008} YbRh$_2$Si$_2$,\cite{gegenwart_magnetic-field_2002, paschen_hall-effect_2004, friedemann_fermi-surface_2010,kambe_degenerate_2014} CeCoIn$_{5-x}$Sn$_x$ and the group of Ce$M$In$_5$ $(M={\rm {Ir,\ Rh,\ Ru}})$\cite{hegger_pressure-induced_2000, petrovic_heavy-fermion_2001, petrovic_new_2001, bauer_superconductivity_2005, donath_dimensional_2008, hu_non-fermi_2013, choi_temperature-dependent_2012} and many others with different structures.\cite{tokiwa_quantum_2011, slebarski_universal_2005}

Chemical composition change is a well known technique to tune physical properties of many strongly correlated systems. Recently we have searched for novel behavior in Ce-based systems with tetragonal body-centered ThCr$_2$Si$_2$-type structure\cite{tolinski_competing_2013,tolinski_influence_2017} or with hexagonal CaCu$_5$-type structure.\cite{synoradzki_effective_2012, synoradzki_x-ray_2014, synoradzki_magnetic_2015, synoradzki_spin_2016} Increased interest of the scientists is brought by the non-centrosymmetric structures, as it is common for them to exhibit superconductivity.\cite{smidman_superconductivity_2017} The tetragonal structure BaNiSn$_3$ ($I4mm$ space group) forms in Ce-based 1-1-3 silicides, such as CeRhSi$_3$,\cite{kimura_pressure-induced_2005, kimura_extremely_2007} CeIrSi$_3$\cite{okuda_magnetic_2007} or CePtSi$_3$\cite{kawai_magnetic_2007} and germanides, such as CeIrGe$_3$\cite{muro_contrasting_1998-1, honda_pressure-induced_2010} or CeRhGe$_3$.\cite{muro_contrasting_1998-1, hillier_muon_2012} The aim of the present studies is to use the chemical substitution in order to provide a wide evidence of the QCP existence and anomalous behaviors around it in the antiferromagnetic CeCoGe$_3$ diluted with a paramagnetic CeFeGe$_3$ compound. CeCoGe$_3$ has three antiferromagnetic phase transitions at $T_{\rm N1} = 21 \ {\rm K}$, $T_{\rm N2} = 12 \ {\rm K}$, and $T_{\rm N3} = 8 \ {\rm K}$. \cite{pecharsky_unusual_1993,thamizhavel_unique_2005,budko_magnetoresistance_1999,kaneko_multi-step_2009,smidman_neutron_2013} Moreover, it shows superconductivity under hydrostatic pressure.\cite{kawai_pressure_2007,settai_pressure-induced_2007} Adding a small amount of Si in the place of Ge leads to even a more complex magnetic structures. \cite{krishnamurthy_non-fermi-liquid_2002, kanai_observation_1999, larrea_j._high_2013} CeFeGe$_3$ is a paramagnet with a high Kondo temperature (over $100 \ {\rm K}$) \cite{yamamoto_new_1994,yamamoto_cefege_1995}. Substituting cobalt with iron and creating the CeCo$_{1-x}$Fe$_{x}$Ge$_3$ alloys series is expected to compensate magnetic ordering of CeCoGe$_3$ at QCP. It was previously reported that for $x \sim 0.6$ a neighborhood of QCP can be attained.\cite{de_medeiros_phase_2001} In this paper the magnetic susceptibility, specific heat, resistivity, magnetoresistance and thermoelectric power measurements have been carried out for series of samples with $0\leq x \leq 1$. 

\section{\label{exper}Experimental}

Polycrystalline samples were prepared of the stoichiometric amounts of high purity elements of Ce $(99.9\%)$, Co $(99.99\%)$, and Ge $(99.999\%)$ and synthesized in the induction furnace in an argon atmosphere. Turning upside-down and remelting several times was applied to ensure the homogeneity. The samples were also annealed in the Ta foil in quartz tubes at the temperature of $750^\circ {\rm C}$ for $120 \ {\rm h}$. For transport measurements the samples were cut to proper dimensions and shape by diamond and wire saws. 
To verify the samples structure the X-ray diffraction (XRD) measurements were carried out in room temperature using D2 PHASER Bruker device. The data were refined using the FULLPROF program, which has shown and confirmed that all the studied compounds are isostructural, single phase, and crystallized in the desired tetragonal BaNiSn$_3$-type structure. 
Further characterization of the samples was performed using the measurements options of the Quantum Design Physical Property Measurement System (QD-PPMS). The magnetic susceptibility was measured with usage of the VSM device in the temperature range $2-400 \ {\rm K}$ with magnetic field $B=0.1 \ {\rm T}$ and hysteresis loops were measured at $2 \ {\rm K}$ with magnetic field up to $9 \ {\rm T}$. The specific heat data were collected by the adiabatic heat pulse method in the temperature range $1.9-295 \ {\rm K}$ for zero magnetic field and with an applied magnetic field for the temperature range $1.9-25 \ {\rm K}$. 
To measure the temperature dependence of resistivity in the temperature range $1.9-300 \ {\rm K}$ a four probe method was used. The magnetoresistance measurements were performed in the temperature range $2-30 \ {\rm K}$ with magnetic field values up to $9 \ {\rm T}$. The thermoelectric power and the thermal conductivity studies were carried out with the thermal transport option (TTO) using a four probe mode.

\section{\label{results}Experimental results}

To identify unambiguously the critical lines between various regions of the Doniach diagram of CeCo$_{1-x}$Fe$_{x}$Ge$_3$, we managed to extract the characteristic temperatures $T^*$, employing different complementary methods like magnetic susceptibility, specific heat, electrical resistivity, magnetoresistance and thermoelectric power. In the following subsections the analysis of these results is presented and finally the magnetic phase diagram is constructed. Simultaneously, the manifestations of the non-Fermi liquid behavior have been searched in the neighborhood of QCP. 

\subsection{\label{mag}Magnetic properties}
\begin{figure}[h]
\centering
\includegraphics[width = 8.6cm]{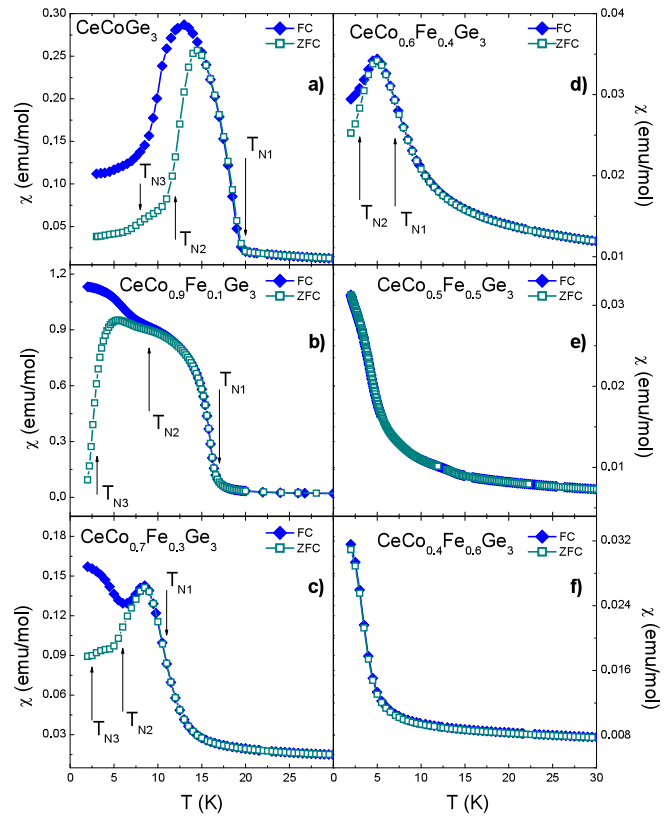}
\caption{\label{fig1} ZFC and FC curves of the magnetic susceptibility for the CeCo$_{1-x}$Fe$_{x}$Ge$_3$ samples at $B = 0.1 \ {\rm T}$.}
\end{figure}

Low temperature magnetic susceptibility for all prepared samples is presented in Fig. \ref{fig1}. Previously, it has been found that a single crystalline CeCoGe$_3$ shows three phase transitions, at $T_{\rm N1} = 21 \ {\rm K}$, $T_{\rm N2} = 12 \ {\rm K}$, and $T_{\rm N3} = 8 \ {\rm K}$,\cite{thamizhavel_unique_2005} whereas for polycrystalline compound $T_{\rm N3}$ has not been observed.\cite{pecharsky_unusual_1993} In Fig. \ref{fig1}(a) there is a clear transition at $T_{\rm N1} = 21 \ {\rm K}$ and two inflections at $T_{\rm N2} = 12 \ {\rm K}$ and $T_{\rm N3} = 8 \ {\rm K}$. Particular transitions are still visible for samples with $x = 0.1, \ 0.3$ and 0.4 with the transition at $T_{\rm N1}$ shifting progressively to lower temperatures. It is visible in Fig. \ref{fig1}(b) that for sample $x = 0.1$ the susceptibility values are roughly four times higher than for CeCoGe$_3$ and the peak of the susceptibility is significantly broadened with less visible transitions $T_{\rm N2}  = 9 \ {\rm K}$ and $T_{\rm N3}  = 3 \ {\rm K}$. This behavior suggests an increased disorder for this level of doping and it is reproducible for another independently prepared sample with $x = 0.1$. Magnetic phase transitions at $T_{\rm N1}$, $T_{\rm N2}$ and $T_{\rm N3}$ are present for  $x = 0.3$ [Fig. \ref{fig1}(c)] with temperatures  $11 \ {\rm K}$, $6 \ {\rm K}$ and $2 \ {\rm K}$, respectively, whereas for $x = 0.4$ only a softening of the transitions at $T_{\rm N1}  = 7 \ {\rm K}$ and $T_{\rm N2}  = 3 \ {\rm K}$ can be detected. For samples $0.5\leq x \leq 0.8$ the phase transitions are not visible for the examined temperature range, indicating a paramagnetic behavior. Concerning the magnetic susceptibility values, in the entire range $0 \leq x\leq 1$ they exhibit a general tendency of decreasing with the growing amount of the Fe addition. 
For samples with $x$ up to 0.4 a split of the ZFC and FC curves is well developed. An exception are the alloys with $x = 0.5,\ 0.6$, and 0.7, which show  almost no difference between ZFC and FC susceptibilities. This can be attributed to a well-ordered crystal structure and the lack of magnetic phase transitions. This indicates that around $x = 0.5$ one can expect the system CeCo$_{1-x}$Fe$_{x}$Ge$_3$ to exhibit the QCP. 

\begin{figure}[h]
\centering
\includegraphics[width = 8.6cm]{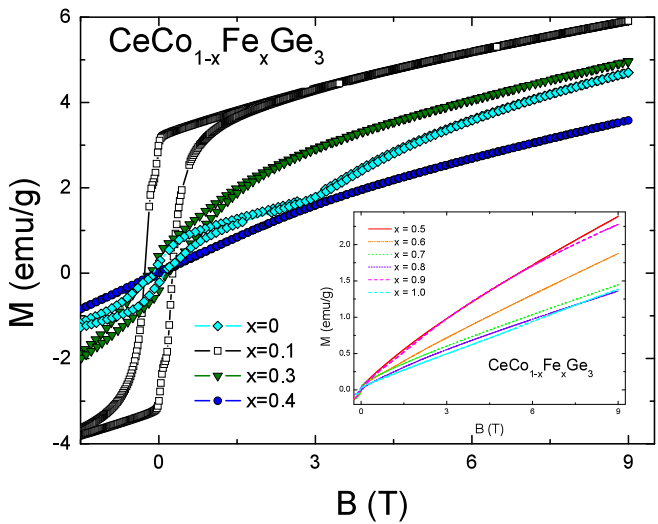}
\caption{\label{fig2}First quarter of the hysteresis loops for all the studied samples at $T= 2\ {\rm K}$. Inset presents results for samples $x\geq 0.5$.}
\end{figure}

Figure \ref{fig2} presents first quarter of the hysteresis loops at temperature of $2 \ {\rm K}$. The polycrystalline CeCoGe$_3$ performs a similar course as was previously measured and described for the single crystal case.\cite{thamizhavel_unique_2005} On the other hand, sample with $x = 0.1$ shows a ferromagnetic character of its hysteresis loop. For $x = 0.3$ this type of isothermal magnetization is also visible. Samples for $ x \geq  0.4$ present paramagnetic character with decreasing values of magnetization when growing the amount of Fe.

The fitting of the Curie-Weiss law to the inverse susceptibility was performed with the expression: 
\begin{equation}
\chi(T)=\frac{N_{\rm A}\mu_{eff}^2}{3k_{\rm B}(T-\theta_p)}+\chi_0
\label{eq1},
\end{equation}
where $N_{\rm A}-$the Avogadro’s number, $\mu_{eff}-$the effective magnetic moment, $k_{\rm B}-$the Boltzmann constant, and $\chi_0-$the temperature independent magnetic susceptibility. For all the samples the effective magnetic moment is around $2.5\ \mu_{\rm B}$, which is comparable to the theoretical value of the free ${\rm Ce}^{3+}$ ion $(2.54\ {\rm \mu_B}$) and this is in agreement with the case of CeCoGe$_3$ \cite{pecharsky_unusual_1993,thamizhavel_unique_2005} and CeFeGe$_3$,\cite{yamamoto_cefege_1995} where it was shown that the contribution of the $3d$ elements to the magnetic moment is negligible. The paramagnetic Curie temperature is negative for all samples, suggesting antiferromagnetic interactions. For $x \leq 0.4$ $\theta_{p}$ is around $-60\ {\rm K}$, which is in agreement with values reported for the polycrystalline sample of CeCoGe$_3$,\cite{pecharsky_unusual_1993} therefore it suggests an antiferromagnetic order for $x \leq 0.4$. For $x = 0.5$, $\theta_{p}$ jumps to almost  $-100\ {\rm K}$, which we ascribe to the appearance of the Kondo interactions. With further increase of the Fe content, the paramagnetic Curie temperature changes to about  $-140\ {\rm K}$.

The magnetic susceptibility is also a useful tool to indicate if NFL exists in the system studied. In the NFL area one usually observes a temperature dependence like $\chi \sim1-(T/T_{0})^{1/2}$ or $\chi \sim \chi_{0}-{\rm ln}(T/T_{0})$.\cite{brian_maple_non-fermi_2010} For the samples $x = 0.5, \ 0.6$ and 0.7 there is a noticeable increase of the magnetic susceptibility values with a decrease of temperature. Therefore, a square root dependence has been used to fit the data for the lowest temperatures. It provides a clear evidence that the NFL behavior may be present for samples with $x = 0.5,\ 0.6$, and 0.7, and therefore CeCo$_{1-x}$Fe$_{x}$Ge$_3$ can be close to QCP for $x \sim 0.5$.  

\begin{figure}[h]
\centering
\includegraphics[width = 8.6cm]{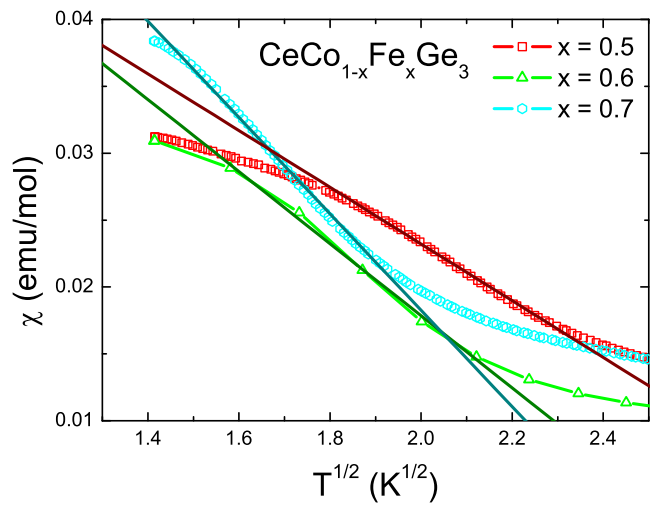}
\caption{\label{fig3}Magnetic susceptibility at low temperatures fitted according to the dependence $\chi \sim1-(T/T_{0})^{1/2}$ suggesting NFL behavior for $x = 0.5,\ 0.6,\ 0.7$.}
\end{figure}

\subsection{\label{heat}Specific heat}
\begin{figure}[h]
\centering
\includegraphics[width = 8.6cm]{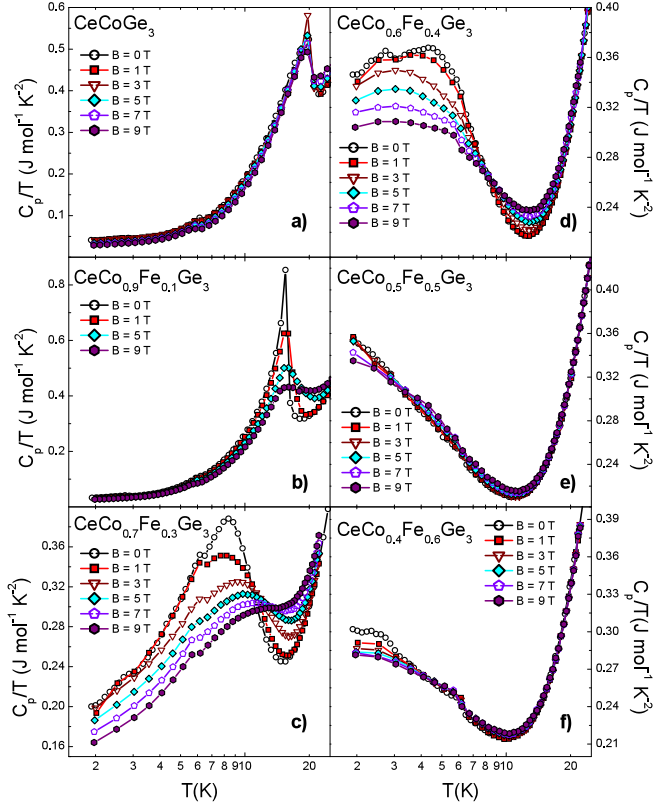}
\caption{\label{fig4}$C_{p}/T\ {\rm vs\ log}T$ results for the CeCo$_{1-x}$Fe$_{x}$Ge$_3$ samples.}
\end{figure}

All the studied samples follow the Dulong$-$Petit law, $C_p= 3NR$, where $N$ is the number of atoms per formula unit, and $R$ is the gas constant. Low temperature $1.9-25 \ {\rm K}$ dependences of the specific heat for different applied magnetic fields are presented in Fig. \ref{fig4}. CeCoGe$_3$ shows a peak around $21\ {\rm K}$, which corresponds to $T_{\rm N1}$ in agreement with previous reports for the polycrystalline and single crystal case.\cite{pecharsky_unusual_1993,thamizhavel_unique_2005} Magnetic field slightly moves the transition to lower temperatures [see inset of Fig. \ref{fig4}(a)], which is known as a possible sign of the antiferromagnetic-type ordering. The sample with $x = 0.1$ shows a sharp phase transition with a large value of the maximum compared to the other samples. The increased amplitude is also visible in the case of the magnetic susceptibility Fig. \ref{fig1}(b)]. A sharp peak in magnetic susceptibility for $x = 0.1$ can be also noticed in the case of the Medeiros results.\cite{de_medeiros_phase_2001} The peak in $C_p/T$ shifts slightly towards higher temperatures with the increase of the magnetic field [Fig. \ref{fig4}(b)] suggesting a ferromagnetic ordering. Similar influence of the magnetic field is observed for $x = 0.3$ and both $x=0.1$ and $x=0.3$ exhibit hysteresis of $M(T)$ in Fig. \ref{fig2}(a) also indicating a ferromagnetic ordering. Two phase transitions are present, $T_{\rm N1} = 11 \ {\rm K}$ and $T_{\rm N2} = 3 \ {\rm K}$ at $0\ {\rm T}$, what is in good agreement with magnetic susceptibility results. These observations clearly indicate that incorporating Fe atoms to the primary AFM compound (at $T_{\rm N1}$) leads to a dramatic change of the magnetic interactions. Due to the disorder developed by the random occupation of the sites with Fe only the main transition at $T_{\rm N1}$ persists but the AFM order evolves probably to a weak noncollinearferrimagnetic order, that is why a small magnetic field is enough to rotate the magnetic moments to a parallel alignment. In consequence one observes a shift of the specific heat peak towards larger temperatures with the growing magnetic field.

Figure \ref{fig4}(e) shows the magnetic field dependence of $C_{p}/T\ {\rm vs}\ T$ at low temperatures for $x = 0.5$. Below $10\ {\rm K}$ clearly a logarithmic growth is visible for $B=0\ {\rm T}$. It can be ascribed to the NFL behavior. The application of a magnetic field over $5\ {\rm T}$ slightly spoils the logarithmic behavior, which is expected because magnetic field is known as a tool to control the system properties around QCP, therefore it can shift a system towards or away from the QCP. A similar $C_p/T \sim -{\rm ln}(T/T_0)$ growth has been observed for $x =0.6$ and $x =0.7$, therefore we conclude that for the substitution region $0.5 \leq x \leq 0.7$ the CeCo$_{1-x}$Fe$_{x}$Ge$_3$ series performs a NFL behavior driven by the QCP.

It is very interesting to verify how the electronic specific heat coefficient $\gamma$ changes with the Fe substitution $x$. Due to the peaks developed by the magnetic ordering for $x \leq 0.5$ and the anomalous growth in the QCP region we analyse the behavior of two characteristic parameters $\gamma$ and $\gamma$$^*$, corresponding to a different temperature regions. The first one is derived from the standard low temperature limit of the Debye model:
\begin{equation}
C_p(T)/T=\gamma+\beta T^2
\label{eq2}.
\end{equation}
The fitting procedure has been done for $25\ {\rm K} <  T < 50\ {\rm K}$, i.e. above the region of phase transitions. For a reasonable comparison this has been consequently kept for the paramagnetic samples. The other parameter, $\gamma^*$, has been determined by extrapolation of $C_{p}/T$ to $T=0$ in the region of the lowest temperatures. Especially, considering the existance of the QCP and NFL behavior for samples with $x = 0.5,\ 0.6,\ 0.7$, the Kondo temperature $T_{\rm K}$ can be derived from the equation \cite{brian_maple_non-fermi_2010}:
\begin{equation}
C_p(T)/T=\gamma^*-[1/T_0\ln(T/T_0)]
\label{eq3},
\end{equation}
where $T_0$ for $f-$electron system is identified as $T_{\rm K}$.\cite{brian_maple_non-fermi_2010} The determined values for the three analysed samples are $10\ {\rm K}$, $18\ {\rm K}$, and $38\ {\rm K}$ for $x = 0.5,\ 0.6$ and 0.7, respectively. $T_{\rm K}$ can be also calculated from the relation \cite{andrei_solution_1983} $T_{\rm K}=0.68R/\gamma^*$ as well as using the paramagnetic Curie temperature $T_{\rm  K}= |\theta_p|/4.5$.\cite{gruner_magnetic_1974} A comparison of the evaluated values of $T_{\rm K}$ is enabled by Table \ref{tab1}. In general, the results are in good agreement with each other. $T_{\rm K}$ rises linearly with the increase of the Fe content, meaning an enhancement of the Kondo interactions.

In Ce$-$based compounds the magnetic, transport and thermodynamic properties can be significantly affected by the excitations between the energy levels of the crystal electric field (CEF). Firstly, non-magnetic contribution of LaCoGe$_3$ and LaFeGe$_3$ has been subtracted with a proper linear combination of these two compounds for particular samples:
\begin{equation}
\begin{split}
&C_{\rm mag}(T)/T[{\rm CeCo}_{1-x}{\rm Fe}_x{\rm Ge}_3]\\
&=C_{p}(T)/T[{\rm CeCo}_{1-x}{\rm Fe}_x{\rm Ge}_3]\\
&-[C_{p}(T)/T[{\rm LaCoGe_3}](1-{x})\\
&+C_{p}(T)/T[{\rm LaFeGe_3}]({x})]
\label{eq4}.
\end{split}
\end{equation}
To estimate the scheme of the CEF energy levels, and especially its evolution with the change of the Fe content, the subtraction results have been analysed with the Schottky contribution to the specific heat:
\begin{equation}
\frac{C_{\rm CEF}}{T}=\frac{R}{T^2}\left[\frac{\sum_{i=0}^{n-1}\Delta_i ^2 e^{-\frac{\Delta_i}{T}}}{\sum_{i=0}^{n-1}e^{-\frac{\Delta_i}{T}}}-\left(\frac{\sum_{i=0}^{n-1}\Delta_i e^{-\frac{\Delta_i}{T}}}{\sum_{i=0}^{n-1}e^{-\frac{\Delta_i}{T}}}\right)^2\right]
\label{eq5}.
\end{equation}
Samples with x $\geq$ 0.5 perform an increase of the magnetic specific heat at low temperatures, which is commonly assigned to a heavy fermion state. It can be described by the Schotte-Schotte formula\cite{schotte_interpretation_1975}:
\begin{equation}
\frac{C_{\rm Kondo}}{T}=\frac{k_{\rm B}N_{\rm A}T_{\rm K}}{\pi T^2}\left[1-\frac{T_{\rm K}}{2\pi T}\Psi'\left(\frac{1}{2}+\frac{T_{\rm K}}{2\pi T}\right) \right]
\label{eq6},
\end{equation}
where $\Psi'$ is the derivative of the Digamma function. By combining the Schottky and the Schotte-Schotte contributions to the specific heat the values of the CEF energy levels could be estimated. $T_0$ is of the order of the Kondo temperature; however, it can differ from other estimations due to the approximations of this model. The results are presented in Table \ref{tab1}. It can be noticed in Table \ref{tab1} that both excited levels do not change much for samples with $x \leq 0.5$, while for $x = 0.6$ a clear decrease for $\Delta$$_1$ and $\Delta$$_2$ levels is observed with a further slight reduction for $x > 0.6$. The occurrence of this change is probably connected with a change of the dominating type of interactions, i.e. the enhancement of the Kondo interactions reflected in the increase of $T_{\rm K}$ with the growing Fe content.
\begin{table*}
\caption{\label{tab1} Values of $T_{\rm K}$ obtained with various methods, $\gamma$, $\gamma^*$ and the calculated CEF energy levels; $T_{\rm K1}$ denotes $T_{\rm K}=0.68R/\gamma^*$, $T_{\rm K2}$ denotes $T_ {\rm K}= |\theta_p|/4.5$.}
\begin{ruledtabular}
\def\arraystretch{1.5}%
\begin{tabular}{ccccccc}
$x$ (Fe) & $T_{\rm K1}\ {\rm (K)}$ & $T_{\rm K2}\ {\rm (K)}$ & $\gamma\ {\rm {(mJ\ mol^{-1}\ K^{-2})}}$ & $\gamma^*\ {\rm {(mJ\ mol^{-1}\ K^{-2})}}$ & $\Delta_1\ {\rm (K)}$ & $\Delta_2\ {\rm (K)}$\\
\hline
0 & - & - & 231 & - & 139 & 311\\
0.1 & - & - & 137 & - & 143 & 288\\
0.3 & - & - & 119 & - & 138 & 332\\
0.4 & - & - & 109 & - & 119 & 340\\
0.5 & 13 & 20 & 122 & 422 & 120 & 308\\
0.6 & 17 & 26 & 131 & 342 & 81 & 225\\
0.7 & 23 & 34 & 124 & 245 & 80 & 200\\
0.8 & 28 & 36 & 124 & 199 & 75 & 190\\
0.9 & 27 & 22 & 102 & 209 & 68 & 181\\
1.0 & 45 & 32 & 85 & 126 & 73 & 176\\
\end{tabular}
\end{ruledtabular}
\end{table*}
\subsection{\label{res}Resistivity}

\begin{figure}[h]
\centering
\includegraphics[width = 8.6cm]{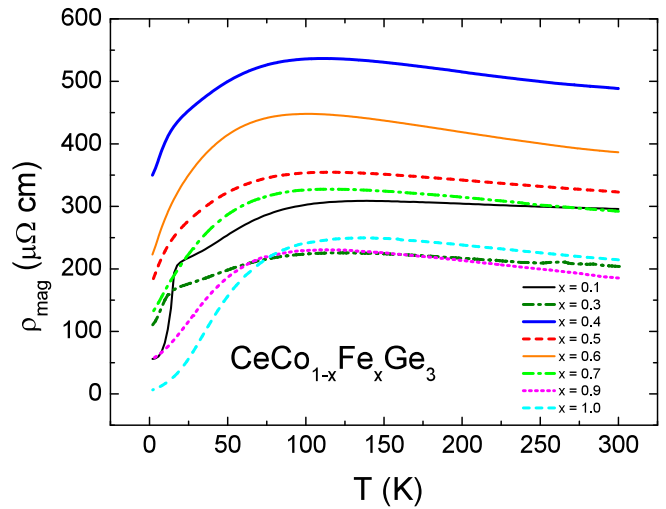}
\caption{\label{fig5}Magnetic part of the resistivity for series CeCo$_{1-x}$Fe$_x$Ge$_3$.}
\end{figure}

Resistivity results present visible anomalies for $x = 0.1,\ 0.3$ and 0.4, that indicate phase transitions at similar temperatures as it was deduced from magnetic susceptibility and specific heat measurements, while the rest of the samples present paramagnetic character [Fig. \ref{fig5}]. The residual resistivity $\rho_0$ varies within the range $7.5-209\ {\rm \mu \Omega cm}$, with the value of $7.5\ {\rm \mu \Omega cm}$ reached for the CeFeGe$_3$ sample. The detailed results are included in Table \ref{tab2}. Large $\rho_0$ values in middle level of the Fe doping may indicate an increased disorder on the $3d$ site. However, the trend of increasing $\rho_0$ values towards the concentration of Fe with $x = 0.5-0.6$ can be also explained by the heavy fermions nature, which is connected with a neighborhood of the QCP. It can be noticed in Table \ref{tab1} that $\gamma^*$ is enhanced in this concentration range. 
The effect of the increased $\rho_0$ is diminishing with the growing addition of Fe as the system is moving out of the vicinity of the QCP.
Using reported previously results for LaCoGe$_3$ and LaFeGe,$_3$ \cite{pecharsky_unusual_1993,yamamoto_cefege_1995} the nonmagnetic contribution has been subtracted to obtain the magnetic part of the CeCo$_{1-x}$Fe$_{x}$Ge$_3$ resistivities. The $\rho (T)$ dependences of the lanthanum-based samples were fitted with the Bloch-Gr{\"u}neisen formula:
\begin{equation}
\rho(T)=\rho_0+4RT\left(\frac{T}{\Theta_{\rm D}}\right)^4\int_0^{\frac{\Theta_{\rm D}}{T}}\frac{x^5 dx}{\left(e^x-1\right)\left(1-e^x\right)}
\label{eq7}.
\end{equation}
The magnetic part of resistivity of CeCo$_{1-x}$Fe$_{x}$Ge$_3$ has been extracted from the following combination of the nonmagnetic reference compounds:
\begin{equation}
\begin{split}
&\rho_{\rm mag}(T)[{\rm CeCo}_{1-x}{\rm Fe}_x{\rm Ge}_3]\\
&=\rho(T)[{\rm CeCo}_{1-x}{\rm Fe}_x{\rm Ge}_3]\\
&-[\rho(T)[{\rm LaCoGe_3}](1-{x})\\
&+\rho(T)[{\rm LaFeGe_3}]({x})].
\label{eq8}
\end{split}
\end{equation}
All samples exhibit a wide peak around $100\ {\rm K}$ being a result of the lower energy CEF excitation with a possible influence of the Kondo effect [Fig. \ref{fig5}]. There is a visible shift of the peak towards higher temperatures for $x > 0.6$, which results from the growing Kondo temperature as well as the related coherence temperature.

Magnetic susceptibility and specific heat measurements delivered information on NFL behavior for samples with $x = 0.5,\ 0.6$ and 0.7. To verify it additionally by the resistivity investigations the power law function has been applied at the lowest temperatures for samples with $x \geq 0.5$:
\begin{equation}
\rho(T)=\rho_0+aT^\alpha
\label{eq9}.
\end{equation}
For NFL it is common that $\alpha <2$, whereas for the QCP neighborhood $\alpha \sim 1$. We find $\alpha =0.90$ and $\alpha =1.03$ for $x$ equal to 0.5 and 0.6, respectively, which suggests the vicinity of the QCP similarly as it was reported previously.\cite{de_medeiros_phase_2001, skokowski_magnetoresistance_2017} Values of the power parameter are increasing with the increase of the Fe concentration, reaching almost 2 for CeFeGe$_3$.
\begin{table*}
\caption{\label{tab2} Parameters extracted from the analysis of the resistivity and thermoelectric power results.}
\begin{ruledtabular}
\def\arraystretch{1.5}%
\begin{tabular}{ccccccc}
$x$ (Fe) & $T_{\rm K}\ {\rm (K)}$ & $T_{\rm CEF}\ (N_f=2)\ {\rm (K)}$ & $T_{\rm CEF}\ (N_f=4)\ {\rm (K)}$ & $T_{\rm CEF}\ (N_f=6)\ {\rm (K)}$ & $\rho_0\ {\rm (\mu \Omega\ cm)}$ & $\alpha$\\
\hline
0.1 & - & 108 & 215 & 322 & 56 & -\\
0.3 & - & 110 & 219 & 329 & 105 & -\\
0.4 & - & 107 & 213 & 320 & - & -\\
0.5 & - & 94 & 187 & 281 & 166 & 0.90\\
0.6 & 60 & 89 & 178 & 268 & 209 & 1.03\\
0.7 & 48 & 85 & 170 & 255 & 125 & 1.20\\
0.8 & 52 & 89 & 178 & 268 & 90 & 1.54\\
0.9 & 28 & 89 & 178 & 268 & 57 & 1.72\\
1.0 & 66 & 88 & 173 & 260 & 7.5 & 1.84\\
\end{tabular}
\end{ruledtabular}
\end{table*}
\subsection{\label{MR}Magnetoresistance}
\begin{figure}[h]
\centering
\includegraphics[width=8.6cm]{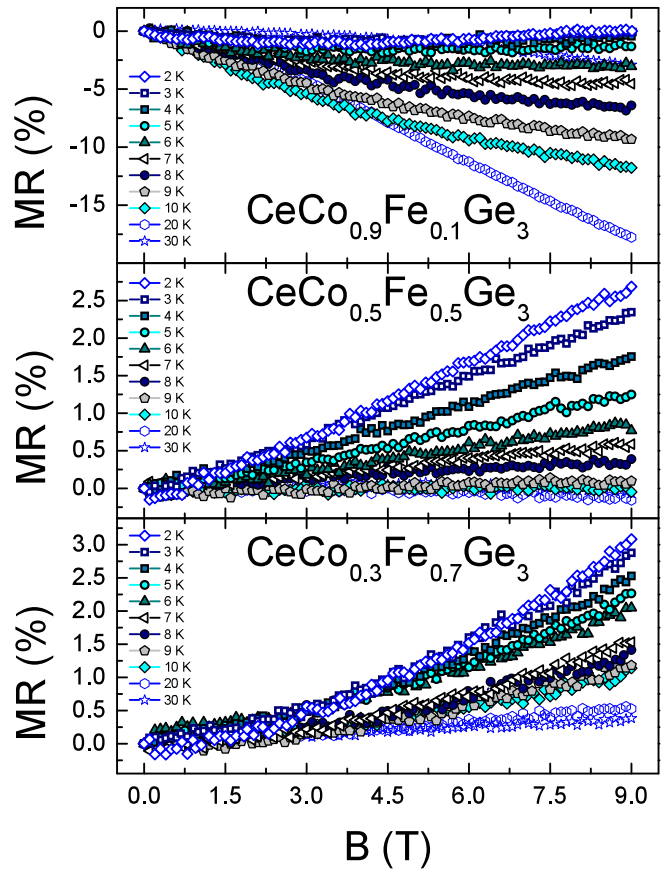}
\caption{\label{fig6}The magnetic field dependence of magnetoresistance isotherms measured at $2-30\ {\rm K}$ range for $x = 0.1,\ 0.5,\ 0.7$ and 0.9.}
\end{figure}

A complementary information on the system studied can be expected by studying the influence of the external magnetic field on the electrical resistivity. Therefore, we have measured the magnetoresistance $(MR)$ according to the formula: $MR=[\rho(H,T)-\rho(0,T)]/\rho(0,T)$. In Fig. \ref{fig6}(a) $MR$ curves for $ x = 0.1$ show small values for the lowest temperatures $(T = 2,\ 3,\ 4\ {\rm K})$, as well as an interesting shape with decrease and upturn for higher magnetic field values. This behavior was reported \cite{nigam_ferromagnetic--antiferromagnetic_1994} as characteristic for spin-glass state, as a competition between ferro- and antiferromagnetic interactions, while small values are connected with spin-disorder scattering. With increase of the temperature the curves are forming known shape for ferromagnets. Those results are in good agreement with conclusions derived from previous sections of this work. For the lowest temperatures for $x = 0.1$ an anomaly is observed with negative sign of $MR$. We ascribe it to the impurity Kondo effect like in the case of the magnetic susceptibility and specific heat results. It is an expected behaviour as CeFeGe$_3$ is a Kondo system.
Previously, we reported for samples $x = 0.3,\ 0.4$ and 0.6 \cite{skokowski_magnetoresistance_2017} that the two first of the mentioned samples show an antiferromagnetic character of $MR$ below $10\ {\rm K}$ with Kondo-like behavior for $20\ {\rm K}$ and $30\ {\rm K}$ , whereas a radical change of the behavior was observed for $x = 0.6$, i.e. NFL type $MR$\cite{nakajima_magnetotransport_2008} appeared below $10\ {\rm K}$ followed by a metallic behavior for $20\ {\rm K}$ and $30\ {\rm K}$. Here, as it is shown in Fig. \ref{fig6}(b), the sample with $x = 0.5$ also presents NFL behavior of magnetoresistance with a close to linear magnetic field dependence, which suggests existence of QCP near this concentration. Sample $x = 0.7$ is similar to $x=0.5$ but with the additional parabolic curvature.
\subsection{\label{TTO}Thermoelectric power}
\begin{figure}[h]
\centering
\includegraphics[width=8.6cm]{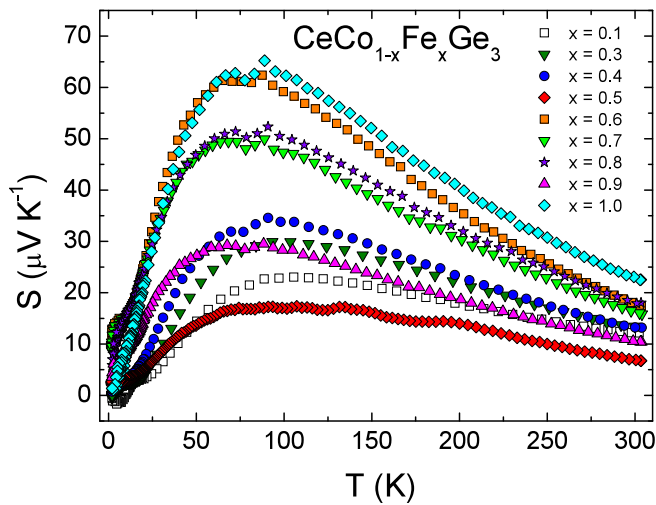}
\caption{\label{fig7}Temperature dependence of the Seebeck coefficient for the CeCo$_{1-x}$Fe$_x$Ge$_3$ series.}
\end{figure}

A unique, sensitive tool to characterize simultaneously the electrical and thermal transport properties is a measurement of the Seebeck coefficient $S$ as a function of temperature. It can reveal features characteristic of CEF, Kondo lattice state, magnetic phase transition as well as NFL behavior. Figure \ref{fig7} shows a common shape known for cerium compounds with a wide maximum around $80\ {\rm K}$. For $x < 0.6$ maximum values of $S$ change with slight tendency of increasing with the Fe content and reaching the highest value for $x = 0.6$, whereas addition of more Fe lowers the maximum values with exception of $x = 1.0$, which has similar values as the sample $x = 0.6$. 
To analyze the $S(T)$ dependences we have used the usual model assuming scattering of electrons from the wide conduction band into a narrow $f-$band approximated by a Lorentzian shape,\cite{gottwick_transport_1985,tolinski_thermoelectric_2010} i.e. the formula reads:
\begin{equation}
S(T)=S_0(T)+S_f(T)=aT+\frac{2}{3}\frac{k_B}{|e|}\frac{\pi^2 E_f T}{\left( \pi^2 / 3\right)T^2+E_f^2+W_f^2}
\label{eq9},
\end{equation}
where the term $S_0(T) = aT$ has been added to account for the usual diffusion contribution. $E_f$ and $W_f$ are the position and width of the $f-$band in Kelvins. Furthermore, it is assumed that $E_f = T_{\rm K}$ and $W_f = \pi T_{\rm CEF}/N_f$ with $T_{\rm CEF}$ being an overall CEF splitting and $N_f$ the orbital degeneracy $2J+1$. The derived values of the fitting parameters are compiled in Table \ref{tab2}. $T_{\rm CEF}$ values fall well in the range determined for $\Delta_1$ and $\Delta_2$ (see Table \ref{tab1}) in the specific heat studies, moreover, the changes with $x$ keep a similar tendency.

\begin{figure}[h]
\centering
\includegraphics[width=8.6cm]{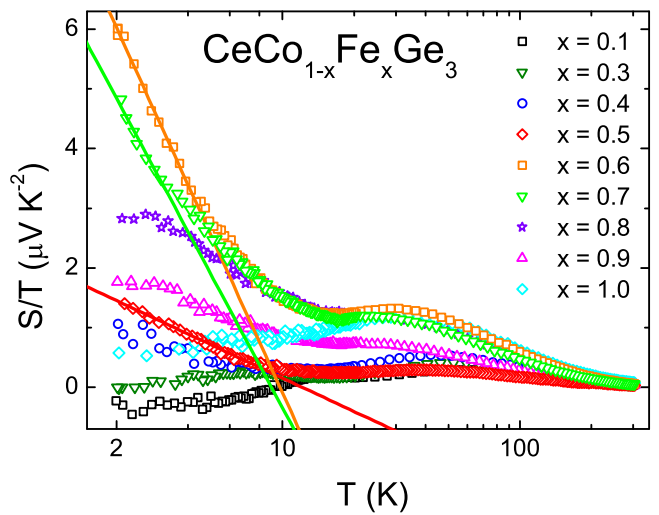}
\caption{\label{fig8}$S/T$ vs log$T$ curves for series CeCo$_{1-x}$Fe$_x$Ge$_3$ in logarithmic scale. Solid lines fitted for $x=0.5,\ 0.6$ and 0.7 follow the relation $S/T\sim -{\rm ln}(T/T_0)$.}
\end{figure}

Low temperatures $S/T$ vs $T$ (Fig.\ref{fig8}) dependences are very similar to $C_p/T$ vs $T$ curves for respective samples. For samples with $x \leq 0.4$ there are anomalies connected with $T_{\rm N}$. On the other side of the system, for $x \geq 0.8$, $S/T$ keeps a constant value, which is known as typical of the Fermi liquid regime. The NFL behavior in the thermoelectric power is usually characterized by the relation $S/T\sim -{\rm ln}(T/T_0)$. In the case of the CeCo$_{1-x}$Fe$_{x}$Ge$_3$ system it is noticeable for the samples $x = 0.5,\ 0.6$, and 0.7. Considering possibilities of SDW QCP or Kondo breakdown QCP as reported by Kim, et al.\cite{kim_thermopower_2010} the signs of N{\'e}el temperature as well as similarity to specific heat results for $x = 0.3$ and 0.4 suggest the SDW scenario.
\section{\label{disc}Discussion}
In this section we will present a few conceptional theories for the whole phase diagram of the  CeCo$_{1-x}$Fe$_{x}$Ge$_3$ system, as well as interpret registered results of NFL behavior in terms of the QCP and the way how it is working in this system.
\subsection{\label{RKKY}Magnetism}
\begin{figure}[h]
\centering
\includegraphics[width=8.6cm]{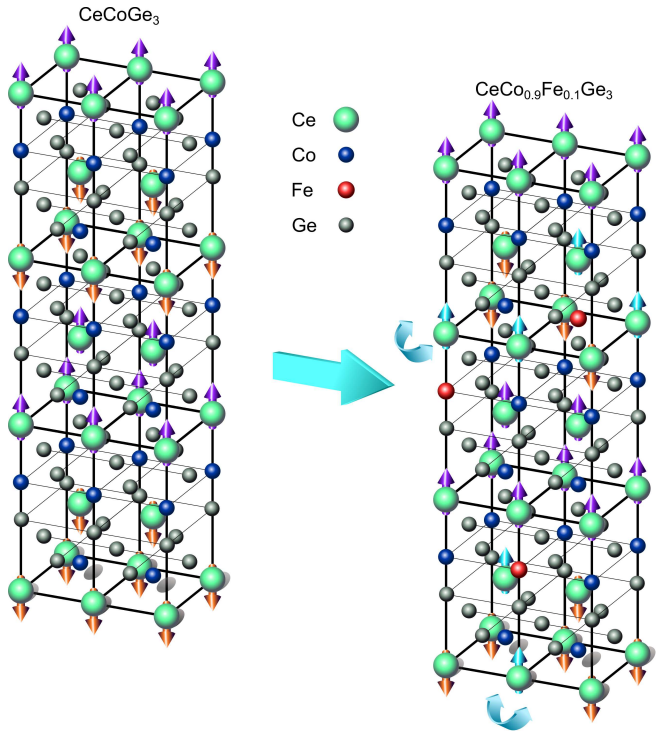}
\caption{\label{fig9}Schematic illustration of the magnetic moments arrangement for CeCo$_{0.9}$Fe$_{0.1}$Ge$_3$ based on magnetic structure at $2\ {\rm K}$ described by Smidman et al.\cite{smidman_neutron_2013} Light blue arrows assign Ce magnetic moments reoriented due to local disorder provided by the Fe doping.}
\end{figure}

As has been already noticed, the doping with Fe is critical for the magnetic structure of the parent CeCoGe$_3$ compound. The already well established antiferromagnetic ordering (including the secondary low temperature phase transitions) is disturbed by the impurity Fe atoms, probably by modification of the RKKY interaction between the Ce ions. Effectively, a non-collinear ferromagnetic type of ordering can be suspected for $x \geq 0.1$. For $x$ around 0.1 the Fe atoms cannot fill regularly all the unit cells. As in the primitive cell there are 5 atoms of $3d$ elements at position $2a$ (as illustrated schematically in Fig.\ref{fig9}), the minimum doping needed to fill statically every cell is $20\%$ Fe, hence the disorder and broadening of the magnetic transitions are especially enhanced. Nevertheless, the overall tendency is a reduction of the ordering temperature towards $0\ {\rm K}$ and lowering of the magnetic susceptibility (with except of $x = 0.1$) with the increase of the Fe content up to about $x=0.5$. Considering that  Fe has one electron less than Co and the density of electronic states at the Fermi level is lower $(1.54/eV$ and $1.72/eV$ for Fe and Co, respectively), it obviously can weaken the RKKY interaction according to $T_{\rm RKKY} \sim J^{2}N(E_{\rm F})$. For $ x > 0.5$ a paramagnetic state stabilizes with a well-established Kondo temperature $T_{\rm K}$ and the coherence temperature $T_{coh}$. 

\subsection{\label{NFL}Non-Fermi liquid behavior and the quantum critical point}
\begin{figure}[h]
\centering
\includegraphics[width=8.6cm]{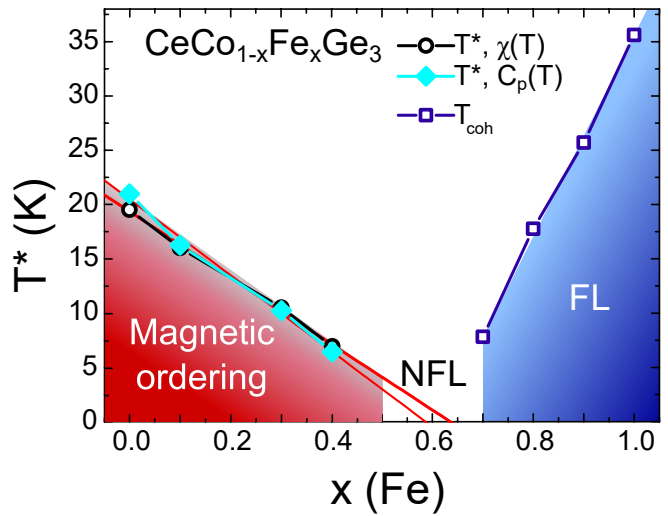}
\caption{\label{fig10}Tentative phase diagram for the CeCo$_{1-x}$Fe$_{x}$Ge$_3$ system. Red lines are linear extrapolations of phase transitions temperatures $T^*$ for both magnetic susceptibility and specific heat results. Phase transition predicted for $x=0.5$ is already not clearly detectable, probably due to the increasing influence of the NFL behavior.}
\end{figure}
NFL behavior has been registered for three samples: $x = 0.5,\ 0.6$ and 0.7 with every experimental method used in frames of this work: magnetic susceptibility performs $-{\rm ln}(T)$ dependence similarly to $C_p/T$ and $S/T$ results, low temperatures resistivity has shown linear temperature dependence and $MR$ revealed sharp change of behaviour compared to lower additions of Fe. It is in good agreement with the Doniach diagram, which predicts that QCP should be around $ x \sim 0.5$ for the discussed series. A tentative phase diagram is displayed in Fig.\ref{fig10}. $T^*$ denotes the temperature of the transition to the magnetically ordered state and corresponds to the values determined from the magnetic susceptibility and specific heat measurements. For Fermi liquid region the coherence temperature\cite{continentino_quantum_2005} has been determined from the resistivity curves assuming that $T_{coh}$ is a threshold below which the dependence $\rho(T) \sim T^2$ starts to work. The extrapolation of $T^*$ in the phase diagram designates $x_c = 0.58$ and 0.65 for the magnetic susceptibility and specific heat methods, respectively, which means that it is needed to adjust the $\delta$ parameter even further with magnetic field or pressure. Linear decrease of a phase transition, as it occurs in the discussed series, was firstly observed for CeCu$_{6-x}$Au$_x$ system.\cite{rosch_mechanism_1997,stockert_two-dimensional_1998} Similarly, results presented in this paper for thermodynamics do not follow the relations $C/T \sim 1-BT^{1/2}$ and $\rho\sim \rho_0 + AT^{3/2}$ typical of the dimensionality $d = 3$ and dynamic exponent $z =2$, therefore a scenerio of two-dimensional fluctuations\cite{millis_effect_1993,varma_quantum_2015} is a good possibility because it conserves $C/T \sim -{\rm ln}(T/T_0)$ and $ \rho \sim \rho_0 + aT$, as it was confirmed previously.\cite{stockert_two-dimensional_1998,bauer_superconductivity_2005} However, at the lowest temperatures it is possible to cross the system to the three-dimensional fluctuations, which can be detected with more subtle methods.\cite{donath_dimensional_2008} Assuming that in CeCo$_{1-x}$Fe$_{x}$Ge$_3$ system this scenario is applicable, planes with magnetic ordering have to be existing. Considering crystal structure and neutron diffraction results,\cite{smidman_neutron_2013} magnetic moments are oriented in direction of the $c$ axis and two planes perpendicular to the $c$ axis are possible: those containing Ce atoms from the edges of the unit cell, or one in the middle of the cell. Taking in account positions of Co and Fe in the unit cell, the orientation of Ce magnetic moments is connected with the distribution of those elements in their positions. Interactions between these two $3d$ elements and Ce ions are different, which is documented with different magnetisms for CeCoGe$_3$ and CeFeGe$_3$ compounds. The interplay of these interactions at equal distribution of the $3d$ elements can lead to magnetic instability in both Ce planes. Therefore, QCP in the CeCo$_{1-x}$Fe$_{x}$Ge$_3$ system is driven not only by the chemical pressure due to the substitution of $3d$ element of different radius, but the density of states at the Fermi level plays also an important role.

\section{Conclusion}
We have shown that the CeCo$_{1-x}$Fe$_{x}$Ge$_3$ system provides a unique possibility to study interplay of magnetic order, Kondo scattering, local disorder, and electronic structure by a transformation between two isostructural compounds: antiferromagnetic CeCoGe$_3$ and paramagnetic heavy fermion CeFeGe$_3$. The phase transitions observed for the polycrystalline sample agree perfectly with previous results for a single crystal. By progressive substitution of Fe in place of Co we have found that the magnetic order is critically sensitive to the Fe addition. Nevertheless, the ordering temperature behaves in a monotonic way decreasing down to a region governed by the non-Fermi liquid behavior, which we locate in the range $0.5 \geq x \geq 0.7$. It implies that the quantum critical point exists within this substitution level. The NFL behavior has been corroborated by the observation of adequate power laws or logarithmic behaviors in low temperature dependences of the magnetic susceptibility, specific heat, electrical resistivity, magnetoresistance, and thermoelectric power. It should be emphasized that the transformation CeCoGe$_3 - $CeFeGe$_3$ is not just a chemical pressure, which mimics the real hydrostatic one but the electronic structure modification plays an important role (the transformation is not isoelectronic). It opens interesting area for further, being in progress, studies on the CeCo$_{1-x}$Fe$_{x}$Ge$_3$ system.

\bibliographystyle{apsrev4-1}

\end{document}